\documentclass[newstyle,twocolumn,proceedings]{rmaa}

\usepackage{psfig}

\title{Looking for the First Galaxies with EMIR/GTC}
\author{R. Pello and D. Schaerer
  \affil{Laboratoire d'Astrophysique de l'Observatoire
Midi-Pyr\'en\'ees, Toulouse, France} }

\fulladdresses{ 
\item R. Pell\'o and D. Schaerer: Laboratoire d'Astrophysique de l'Observatoire
Midi-Pyr\'en\'ees, UMR5572, 14 Av. Edouard-B\'elin, F-31400 Toulouse,
France (roser,schaerer@ast.obs-mip.fr).}

\shortauthor{Pello \& Schaerer}
\shorttitle{Looking for the First Galaxies with EMIR/GTC}

\keywords{cosmology: early universe --- galaxies: high redshift --- 
galaxies: evolution --- infrared: galaxies }

\abstract{%
The new Pop III models by Schaerer (2002) have been used to
derive the observed properties of the first galaxies in terms of the
expected magnitudes and colors. The dependence of their properties on the
IMF and upper mass limit for star formation are studied.
The emerging synthetic spectra are used to discuss the implications on
different observational features. Strong emission lines, such as Lyman-alpha
and HeII $\lambda$1640, could easily be
detected with a good S/N with near-IR medium resolution spectrographs,
such as EMIR at GTC. Our simulations aim at exploring possible
observational constraints on the formation epoch of the first stars in
galaxies. }
\resumen{%
Los nuevos modelos de estrellas de Poblaci\'on III de Schaerer (2001)
se han utilizado para derivar las propiedades observacionales de las
primeras galaxias en t\'erminos de magnitudes y colores esperados para
estas fuentes. Hemos estudiado la influencia de la IMF y de la masa l\'
\i mite superior para la formaci\'on estelar en las propiedades de
los objetos. Los espectros sint\'eticos as\' \i\ obtenidos se han
utilizado para discutir la posible observaci\'on de dichas
fuentes. Hemos comprobado que es posible detectar ciertas l\' \i neas
de emisi\'on intensas, tales como Ly~$\alpha$ o HeII $\lambda$1640, con buena
relaci\'on se\~nal/ ruido, utilizando un espectr\'ografo de
resoluci\'on intermedia en el dominio pr\'oximo-IR tal como EMIR en el
GTC. Nuestras simulaciones tienen como objetivo explorar las posibles
restricciones observacionales que podr\'an imponerse en un futuro
sobre la \'epoca de formaci\'on de las primeras estrellas en galaxias. }

\listofauthors{R.~Pello \& D.~Schaerer}
\indexauthor{Pello, R.}
\indexauthor{Schaerer, D.}

\begin{document}

\maketitle

\section{Introduction}
\label{sec:intro}

\begin{figure}
\psfig{file=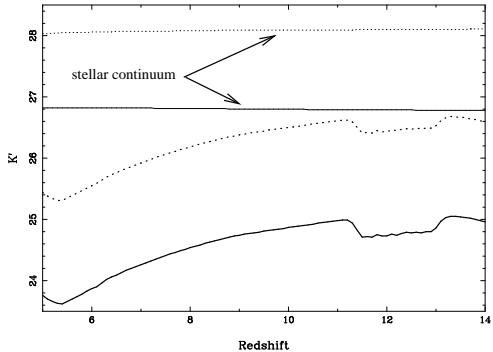,width=7.5cm,angle=270}
  \caption{K' magnitude (Vega system) as a function of redshift over the interval 
z $\sim$  5 to 14, for a starburst model of $10^7$ M$_{\odot}$ and age
$10^4$ yr, with a Salpeter IMF, and stellar masses ranging from 1 to 500 M$_{\odot}$
(dotted lines) and 50 to 500 M$_{\odot}$ (solid lines). 
Pure stellar continuum predictions correspond to the upper thin dotted and 
solid lines, for the two IMF respectively. Lower tick lines display
the predictions for the total spectrum, including lines and nebular 
continuous emission. The two brightness enhancements at z$\sim$ 5.5
and z$\sim$ 11.5-13 correspond to the HeII$\lambda$3203 and 
HeII$\lambda$1640 emission lines entering and crossing this filter.
A typical total magnitude of K' $\sim$ 24-25.5  is expected 
for a stellar halo of $10^7$ M$_{\odot}$.
}
  \label{fig:Kmag}
\end{figure}

   In recent years, important advances have been made
on the modeling of the first stars and galaxies forming out of
primordial matter in the early Universe, the so-called Population 
III objects (cf.\ review of Loeb \& Barkana
2001, proceedings of Weiss et al. 2000 and Umemura \& Susa 2001).
The detection of such sources, which constitute the first building 
blocks of galaxies, remains one of the major challenges of 
present day observational cosmology. Recent modelling efforts
are motivated by the future space facilities such as NGST, which
should  be able to observe these objects at redshifts up to
z$\sim$ 30. Nevertheless, the detection and first studies on the
physical properties of Population III objects could 
likely be started earlier using ground-based 10m class telescopes.
Near-IR multi-object spectrographs, with intermediate resolution 
capabilities, such as EMIR at GTC ($\sim$ 2005) and the future
KMOS at VLT ($\sim$ 2008), will allow observations of Population III
objects up to redshifts of z$\sim$ 18.

   Among the expected direct observational signatures of Pop III stars 
or galaxies (i.e. ensembles/clusters of Pop III stars) we can mention: 


{\bf .} Strong UV emission and characteristic recombination lines of hydrogen 
and He~II, especially Lyman $\alpha$ and HeII (Tumlinson \& Shull 2000, 
Bromm et al. 2001b, Schaerer 2002).

{\bf .} Mid-IR molecular hydrogen lines at 2.12 $\mu$m and longer wavelengths 
formed in cooling shells (Ciardi \& Ferrara 2001).

{\bf .} Individual supernovae whose visibility in the rest-frame optical and 
near-IR could be enhanced due to time dilatation (Miralda-Escude \& Rees 1997,  
Heger et al. 2001).

{\bf .} High energy neutrinos from Pop III gamma-ray bursts eventually
associated with  fast X-ray transients (Schneider et al. 2002).


   Rest-frame UV stellar and nebular continuous and recombination line
emission represent the largest fraction of the energy emitted by
Population III objects, which are generally thought to be
predominantly massive or very massive stars (e.g. Abel et al. 1998,
Bromm et al.  2001a, Nakamura \& Umemura 2001). The predicted rest-frame 
UV to optical spectra of Pop III galaxies including the strongest emission 
lines (Lyman $\alpha$, HeII, etc.) have recently been computed by 
Schaerer (2002). In this paper, we use these synthetic spectra to 
simulate the expected properties of Pop III galaxies, in terms of their colors
and magnitudes, and to study the detectability of the strongest emission
lines. According to our results, efficient photometric
and spectroscopic observations of these features will be possible in the
near-IR domain, thanks to the future imaging and spectroscopic facilities, 
for a large number of sources, thus allowing to derive statistically
significant conclusions about their formation epoch and physical 
properties.

\section{Simulations of PopIII stellar systems} 
\label{sec:simulations}

Simulations have been done to support this scientific case, in
view of the future near-IR facilities, such as EMIR at GTC and 
the future KMOS at VLT (preliminary version: Schaerer \& Pell\'o 2001). 
A popular cosmology is adopted: $\Omega_m$ = 0.3, $\Omega_{\Lambda}$ = 0.7,  
$\Omega_b$=0.05, H$_0$ = 75 km s$^{-1}$Mpc$^{-1}$). The reionization 
redshift is assumed to be $\sim$6, but a small change in this value does not
modify the conclusions of this paper. Lyman series
troughs (Haiman \& Loeb 1999), and Lyman forest following the
prescription of Madau (1995) are included.
We consider a fiducial stellar mass halo of $10^7$ M$_{\odot}$, 
corresponding to a collapsing DM halo of $2 \times 10^8 M_{\odot}$, 
thus typically to $\sim$ 1.5 to 2 $\sigma$ fluctuations between z=5 and 10
(e.g. Loeb \& Barkana 2001).
Magnitudes and S/N ratios are to be rescaled according to this value for
other mass halos. The virial radius is of the order of a few kpc, and thus we
consider that sources are unresolved on a 0.3'' scale, with 
spherical symetry. Simulations accounting
for an extended Lyman~$\alpha$ halo (cf. Loeb \& Rybicki 1999) have also been
performed (Pell\'o \& Schaerer 2002).

\begin{figure}
\psfig{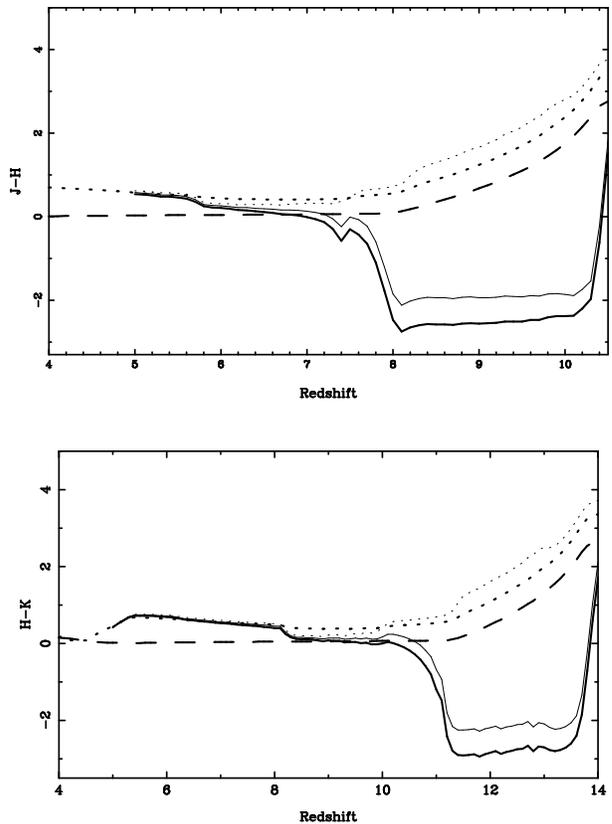}
  \caption{J-H (top) and J-K' (bottom) versus redshift, for 
different fractions of Lyman $\alpha$ emission entering the 
integration aperture: 100\% (thick solid lines), 50\% (thin solid
lines), and 0\% (thin dotted lines). The expected colors obtained within
an integration aperture corresponding to the core (non-resolved)
source, whith Lyman~$\alpha$ emission coming from an extended halo 
(Loeb \& Rybicki 1999), are displayed by thick dotted lines.
The colors expected for a pure
stellar population are given for comparison as dashed thick lines.
In all cases, a Salpeter IMF has been used to derive the synthetic
spectra. 
}
\label{fig:colors}
\end{figure}

   We have considered different IMFs and ages for the stellar population,
as well as two different star formation regimes (starburst and continuous 
star formation), following the prescriptions given by Schaerer (2002).
Magnitudes and colors were derived for these sources in the visible and
near-IR. A complete set of results will be presented in Pell\'o \&
Schaerer (2002). We have also computed the expected S/N ratios for the 
main emission lines. In this case, telescope
parameters correspond to the GTC. The characteristics of the near-IR
spectrograph are set similar to the expectations for EMIR (Balcells et al. 
2000), with 0.2''/pixel, 1'' slit-width, 0.8'' seeing, and a mean total 
efficiency of 40\%. 
All simulations shown subsequently were calculated for a young
population, with a Salpeter IMF from 1 to 500 M$_{\odot}$ (somewhat 
more ``favourable'' than a constant star formation case). Spectral 
resolutions from R=1000 to 5000 were considered.

\section{Summary of results}
\label{sec:results}

\subsection{Photometric properties}

   The importance of the nebular continuous emission, neglected in earlier
studies (Tumlinson \& Shull 2000, Bromm et al. 2001b), is shown in 
Figure~\ref{fig:Kmag}. The predicted magnitude in the K' band is 
typically $\sim$ 24 to 25 for the reference halo mass. Similar effects are seen in J
and H bands (see also Schaerer \& Pello 2001). Examples of broad-band colors
are given in Figure~\ref{fig:colors}. Broad-band colors do not allow 
to constrain physical properties such as the IMF (i.e. the mass 
range of Pop III stars), ages, etc., but could be useful to identify 
the sources on ultra-deep photometric surveys
(cf. below). Spectroscopy 
is needed to study the physics of these objects.

\subsection{Specroscopic properties}

\begin{figure}
\psfig{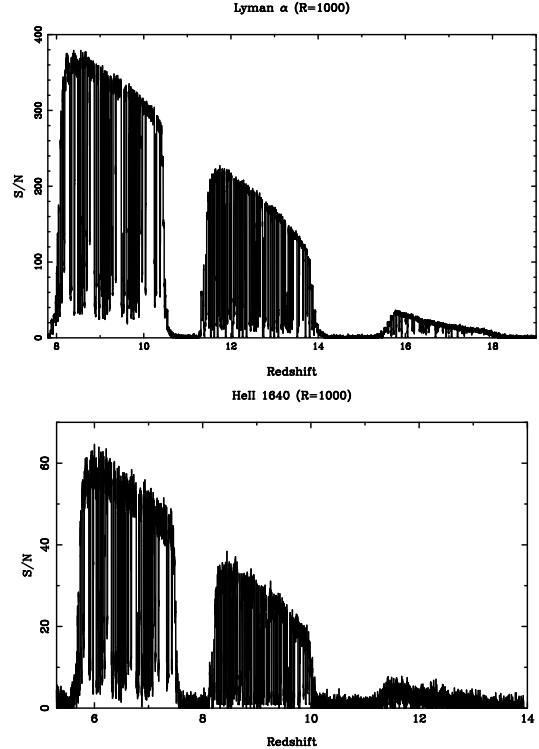}
  \caption{S/N ratio versus redshift for Lyman $\alpha$ (top) and
 HeII $\lambda$1640 (bottom) emission lines, as seen through the JHK bands,
  for a $10^7$ M$_{\odot}$ Pop III stellar halo. The exposure time is
  $10^5$ sec, using a 10m telescope and an EMIR-like spectrograph,
  with an equivalent spectral resolution of R=1000 (rebinned mode). In
  the case of Lyman $\alpha$, the S/N corresponds to 100\% of the energy
  entering the slit, and thus it is the maximum value
  expected for a compact Lyman $\alpha$ halo.  Depending on the extend
  of such halo, the S/N could be lowered up to 5-10\% typically with
  respect to the value shown in this figure.
}
  \label{fig:elines}
\end{figure}

   The expected S/N for the HeII $\lambda$1640 and Lyman~$\alpha$ lines versus the
redshift of the source (observed in the JHK bands), for a spectral
resolution R=1000 and a nominal exposure time of $10^5$ sec, are shown in
Figure~\ref{fig:elines}. The  simulations illustrate the following:

Lyman $\alpha$ can easily be detected with a good S/N over the redshift
intervals z $\sim$ 8 to 18, with some gaps, depending on the spectral
resolution (OH subtraction) and atmospheric transmission. A joint
detection with HeII $\lambda$1640, the strongest HeII line, is possible
for z $\sim$  5.5-7.5 (Lyman $\alpha$ in optical domain), z $\sim$ 8-14 with both lines
in the near-IR, again with some gaps. The typical line fluxes for the
HeII $\lambda$1640 line range between $10^{-17}$ and a few $10^{-18}$
erg/s/cm2. The detection of both HeII $\lambda$1640 and Lyman a allows 
one e.g. to obtain a measure of the hardness of the ionising flux 
which constrains the upper end of the IMF and the age of Pop III 
systems. Measuring the continuum, when possible, will provide 
additional information on the stellar populations, extinction, etc.

  Higher spectral resolution (R $\sim$ 5000) considerably increases the
chances of detection between the sky lines. For unresolved lines (such
as expected for HeII) the resulting decrease of S/N is modest. The
medium spectral resolution is also favoured to attempt to measure the
emission line profiles, in order to distinguish Pop III sources from
potential very high-z AGN (cf.Tumlinson et al. 2001). Once this is
obtained, the data can be rebinned to increase the S/N.

\section{Discussion and future prospects}
\label{sec:discussion}

   The expected number of Pop III objects and primordial QSOs has been
derived by several authors. E.g. in a comprehensive study Ciardi et
al. (2000) show that at z $>$ 8 naked stellar clusters, i.e. objects which
have completely blown out their ISM, and thus avoid local chemical
enrichment, dominate the population of luminous objects. 
Although pilot studies have recently started to explore
the possible formation  of dust in Pop III objects (Todini \& Ferrara
2001), the effect is generally neglected. Based on such assumptions, Oh
et al. (2001) have calculated the predicted number of Pop III objects
detectable in HeII lines with NGST for a one day integration time. 
Their estimate yields between ~ 60 and 4500
sources in a 10'x10' field of view, depending on
the model parameters. The expected density of primordial quasars could
be similar to that of PopIII galaxies (Oh et al. 2001). Thus, 
multiplexing is needed to allow highly efficient observations of relevant
samples of Pop III objects.

   An important issue for spectroscopic studies is the strategy for 
source selection.
Given their peculiar SED (strong nebular continuous emission $+$ lines) 
young Pop III bursts, or Pop III objects with ongoing
massive star formation show distinct characteristics in their near-IR
colors compared to ``normal'' galaxies at any redshift (Pell\'o \&
Schaerer 2002). Such objects could be detected from deep near-IR
photometry based  on a measurement of two colors with accuracies of
the order of ~0.2 mag. Ideal fields for the first studies are lensing
galaxy clusters with areas of strong gravitational amplification, and
other very deep fields with near-IR photometry. Also IFU near-IR
observations could allow to detect the strong HeII emission lines.
Because the
observational signatures of primordial quasars are expected to be
similar to those of genuine Pop III stars, a relatively high resolution 
is needed to obtain line profiles.

   Thanks to its multiplexing, spectral resolution and wide
field-of-view capabilities, EMIR at GTC is the suited instrument to
start exploring the formation epoch of the first stars in galaxies.

\acknowledgements {\bf This work has been done within the framework of the
Cosmos/EMIR collaboration.}
We are grateful to A. Ferrara, J.P. Kneib and J. F. Le Borgne 
for interesting comments and discussion.
Part of this work was supported by the French {\it Centre National de
la Recherche Scientifique}, by the French {\it Programme National de
Cosmologie} (PNC) and {\it Programme National Galaxies}. 


\end{document}